\documentclass[manuscript,screen]{acmart}

\acmConference[Arxiv '20]{Arxiv.org}{November 11, 2020}{Online}
\acmBooktitle{Arxiv}

\AtBeginDocument{%
  \providecommand\BibTeX{{%
    \normalfont B\kern-0.5em{\scshape i\kern-0.25em b}\kern-0.8em\TeX}}}

\usepackage{algorithmic}
\usepackage[flushleft]{threeparttable}
\usepackage{multirow}

\DeclareMathOperator*{\argmax}{argmax}

\usepackage{algorithm2e}
\begin{document}

\title{EM-X-DL: Efficient Cross-Device Deep Learning Side-Channel Attack with Noisy EM Signatures}

\author{Josef Danial}
\affiliation{%
  \institution{Purdue University}
  \city{West Lafayette}
  \state{Indiana}
  \country{USA}}
\email{jdanial@purdue.edu}

\author{Debayan Das}
\affiliation{%
  \institution{Purdue University}
  \city{West Lafayette}
  \state{Indiana}
  \country{USA}}
\email{das60@purdue.edu}

\author{Anupam Golder}
\affiliation{%
  \institution{Georgia Institute of Technology}
  \city{Atlanta}
  \state{Georgia}
  \country{USA}}
\email{anupamgolder@gatech.edu}

\author{Santosh Ghosh}
\affiliation{%
  \institution{Intel Corporation}
  \city{Hillsboro}
  \state{Oregon}
  \country{USA}}
\email{santosh.ghosh@intel.com}

\author{Arijit Raychowdhury}
\affiliation{%
  \institution{Georgia Institute of Technology}
  \city{Atlanta}
  \state{Georgia}
  \country{USA}}
\email{arijit.raychowdhury@ece.gatech.edu}

\author{Shreyas Sen}
\affiliation{%
  \institution{Purdue University}
  \city{West Lafayette}
  \state{Indiana}
  \country{USA}}
\email{shreyas@purdue.edu}

% \thanks{This work was supported in
% part by the National Science Foundation (NSF) under Grant CNS 17-19235, and in part by Intel Corporation.}%
% \thanks{J. Danial, D. Das, and S. Sen are with the School of Electrical and Computer Engineering, Purdue University, West Lafayette, IN, 47906 USA email: (jdanial@purdue.edu, das60@purdue.edu, shreyas@purdue.edu).}%
% \thanks{A. Golder and A. Raychowdhury is with the School of Electrical and Computer Engineering, Georgia Institute of Technology, Atlanta, GA, 30332 USA email: (arijit.raychowdhury@ece.gatech.edu)}%
% \thanks{S. Ghosh is with the Intel Corporation, Hillsboro, OR, 97124, USA, email:(santosh.ghosh@intel.com)}} 

\begin{abstract}
This work presents a Cross-device Deep-Learning based Electromagnetic (EM-X-DL) side-channel analysis (SCA), achieving $>90\%$ single-trace attack accuracy on AES-128, even in the presence of significantly lower signal-to-noise ratio (SNR), compared to the previous works. With an intelligent selection of multiple training devices and proper choice of hyperparameters, the proposed 256-class deep neural network (DNN) can be trained efficiently utilizing pre-processing techniques like PCA, LDA, and FFT on the target encryption engine running on an 8-bit Atmel microcontroller. Finally, an efficient end-to-end SCA leakage detection and attack framework using EM-X-DL demonstrates high confidence of an attacker with $<20$ averaged EM traces.
\end{abstract}

\titlenote{This work was supported in
part by the National Science Foundation (NSF) under Grants CNS 17-19235, CNS 19-35573, and in part by Intel Corporation.}




\maketitle

\section{INTRODUCTION}
With the ever-increasing prevalence of embedded devices and the growth of the Internet of Things (IoT), the security of these devices has become a major concern. Some of the most serious threats to the security of these devices are side-channel analysis (SCA) attacks. By analyzing physical leakage information regarding the power~\cite{kocher_differential_1999}, timing~\cite{kocher_timing_1996}, or electromagnetic (EM) signatures~\cite{agrawal_em_2003}, cryptographic secrets can be extracted. Among the most powerful of these side-channel attacks are profiled attacks~\cite{chari_template_2003}, and recently machine learning (ML) models have been shown to be very effective in this profiled SCA attack scenario using both power and EM measurements~\cite{bartkewitz_efficient_2013,prouff_study_2018}. 
\begin{figure}[!t]
  \centering
  \includegraphics[width=0.8\textwidth]{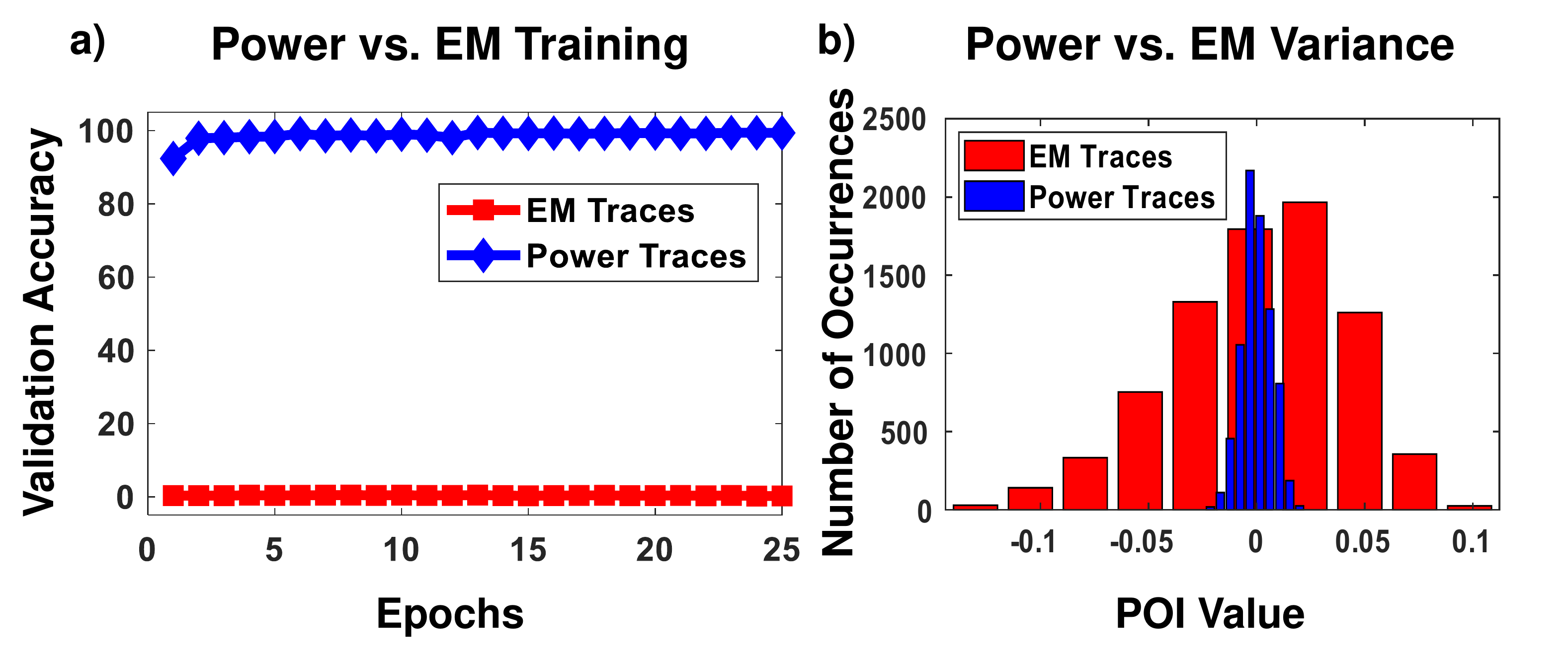}
  \caption{(a) DNN training with high SNR raw power and low SNR raw EM traces. The model learns quickly from the power traces, but is unable to learn from the raw EM traces. (b) The variance of a point of interest (time sample 103) for both power and EM traces, demonstrating significantly lower SNR of the EM traces.}
  \label{power_vs_em}
\end{figure}
\subsection{Motivation}
The main limitation of ML models for profiling SCA attacks is their portability to other target devices. Specifically, these models have been shown to work when the same device is used for both profiling and testing, however, in a real attack, the attacker would use a device to profile, then attack a separate, identical device. This issue of portability has recently been addressed for power ML SCA models on AES-128 in \cite{das_x-deepsca:_2019}, \cite{bhasin_mind_2019}, and also with a 3-class DNN attacking RSA implementations for EM SCA \cite{carbone_deep_2019}. However, these works only consider high SNR scenarios, and with the introduction of SNR reducing countermeasures \cite{das_stellar:_2018} or low cost, low sensitivity EM probes, practical attacks must address the reality of low SNR trace measurements. In this work, we show a deep-learning based cross-device SCA attack with low-SNR EM signatures.
%the first deep-learning based cross-device EM (EM-X-DL) SCA attack on a symmetric key encryption algorithm (AES-128).

\begin{figure}[!t]
  \centering
  \includegraphics[width=0.65\textwidth]{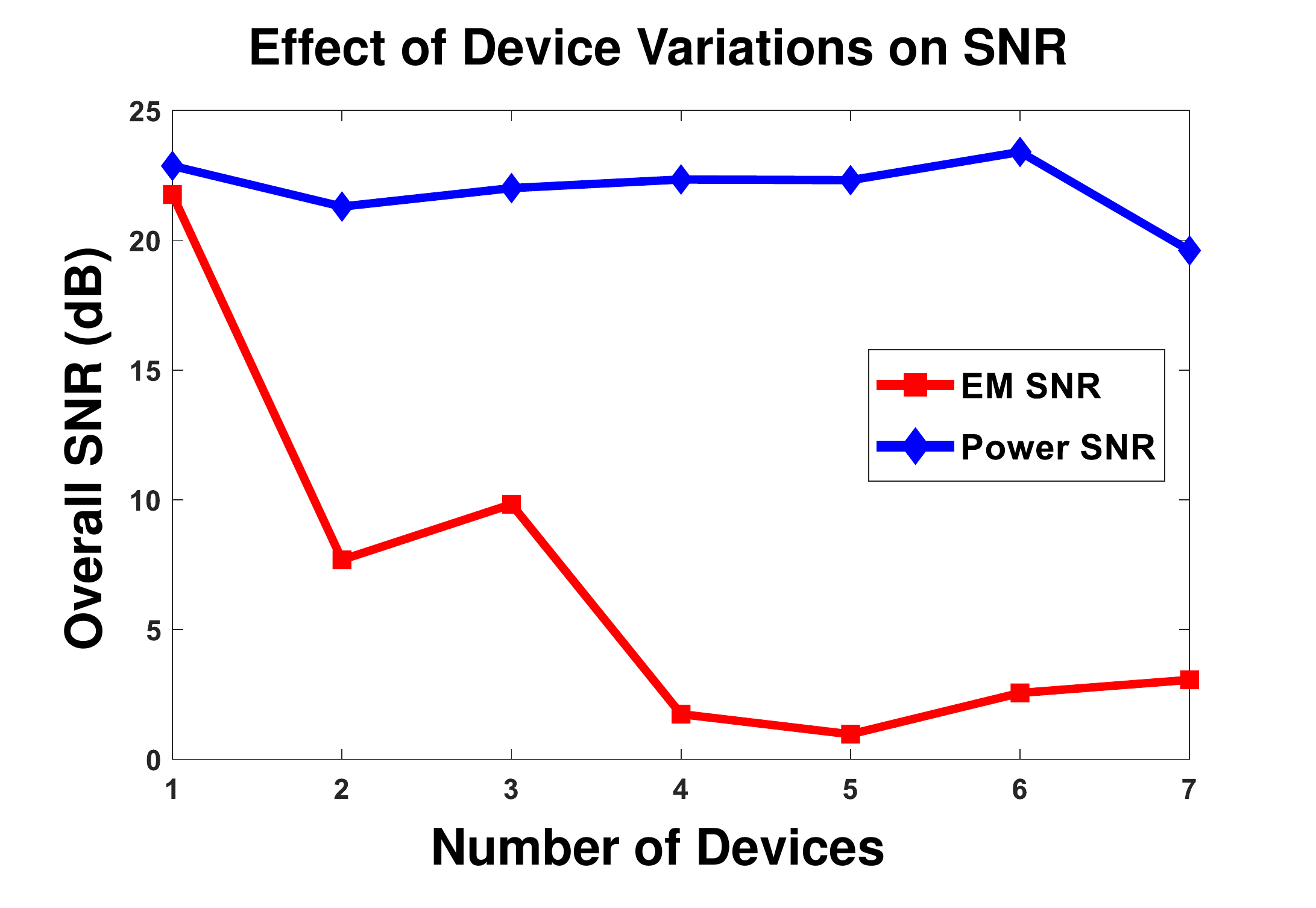}
  \caption{Change in side-channel SNR as devices are added to the dataset for both power and EM. While the power SNR remains fairly high as additional devices are added, the EM SNR drops sharply, indicating a much larger effect of cross-device variations on side-channel leakage in the EM domain compared to power.}
  \label{power_vs_em_snr}
\end{figure}

A 256-class DNN model that can be trained successfully ($>99\%$ validation accuracy)~\cite{das_x-deepsca:_2019} using raw time-domain AES-128 power traces for a particular microcontroller is rendered futile for low SNR EM SCA training even with traces collected from the same device (Figure \ref{power_vs_em}(a)). Figure \ref{power_vs_em}(b) shows the variance of a point of interest (POI, determined using the difference of means approach~\cite{chari_template_2003,choudary_efficient_2018}) across 10K EM and power traces. It clearly shows that the variation in the EM traces is much higher than the power traces, implying significantly lower SNR for the EM signatures. Indeed, when considering the side-channel SNR as defined in~\cite{mangard_hardware_2004} as $\textbf{SNR}=\frac{VAR[Q]}{VAR[N]}$, with $Q$ being the side-channel leakage and $N$ being the noise, there is a large difference when comparing power and EM measurements. The side-channel SNR across a random selection of 7 devices for power traces is 19.6 dB, while the SNR of equivalent EM traces is 3.1 dB, as is seen in Figure \ref{power_vs_em_snr}. Note that the SNR of a single device is comparable for both power and EM, but adding additional devices lowers the SNR drastically, due to the device-to-device variations.  In fact, a majority of the lower SNR in the EM realm is due to \textit{inter-device variations being more prominent in EM compared to power}, again looking at Figure \ref{power_vs_em_snr}. So, to solve the problem of portability, we need to take into account the inter-device variations~\cite{renauld_formal_2011}. To resolve all these issues, we utilize averaging to enhance the SNR, analyze different pre-processing techniques to reduce the dimensionality of the data, and develop an intelligent algorithm to choose the set of training devices for efficient profiling, given the need for more training devices to train a model due to the larger effect of inter-device variations. Finally, we also propose an end-to-end EM-X-DL attack framework to perform EM scanning and find the best point of leakage on an unseen target device. A combination of these techniques allows us to achieve $>90\%$ cross-device accuracy.

\subsection{Contribution}
The specific contributions of this work are:
\begin{itemize}
    \item This work presents a cross-device deep-learning based EM SCA (EM-X-DL) on an AES-128 encryption engine using a 256-class DNN in a low SNR scenario, with ten devices for training and tested on a different set of ten test devices (Sec. 3). 
    \item Effect of different pre-processing techniques including principal component analysis (PCA), linear discriminant analysis (LDA), fast fourier transform (FFT), spectrogram, on handling the portability issue is analyzed and compared, showing that the LDA is the most efficient approach to achieve maximum average cross-device key prediction accuracy of $\sim91.5\%$ with minimum training time (Sec. 3.3).
    \item An algorithm for the optimal selection of the training devices is proposed, so that the number of training devices and thus the overall training time is minimized (Sec. 4, Algo. 1).
    %\item Finally, an end-to-end EM-X-DL attack using the trained DNN model is demonstrated, starting from the EM scanning to finding the point of maximum leakage on the chip, leading to a successful attack at the best location (Sec. 5).
\end{itemize}

\begin{table}[!t]
\centering
\caption{Literature Review for Profiled-Attack Scenario}
\begin{threeparttable}

\begin{tabular}{|c|c|c|c|} 
\hline
\begin{tabular}{@{}c@{}}\textbf{Profiled Attack} \\ \textbf{Scenario}\end{tabular}&
\begin{tabular}{@{}c@{}}\textbf{Measure-} \\ \textbf{ment Type}\end{tabular}&
\begin{tabular}{@{}c@{}}\textbf{Profiling} \\ \textbf{Method}\end{tabular}&
\begin{tabular}{@{}c@{}}\textbf{Corresponding} \\ \textbf{Article}\end{tabular}\\ 
\hline
\multirow{5}{*}{Same-Device} &
\multirow{3}{*}{Power}  & TA & \cite{chari_template_2003}\\ 
\cline{3-4}
                                     && SVM, RF      &                    \cite{bartkewitz_efficient_2013},~\cite{lerman_power_2014} \\ 

\cline{3-4}
                                     && DNN            & \cite{maghrebi_breaking_2016}                               \\
\cline{2-4} &
\multirow{2}{*}{EM}  & TA & \cite{chari_template_2003}\\
\cline{3-4}
                                     && DNN            & \cite{prouff_study_2018}                               \\
\hline
\multirow{7}{*}{\textbf{Cross-Device}} &
\multirow{2}{*}{Power}  & TA & \cite{choudary_efficient_2018}\\ 

\cline{3-4}
                                     && DNN             & \cite{das_x-deepsca:_2019},~\cite{bhasin_mind_2019}                               \\
\cline{2-4} &
\multirow{5}{*}{\textbf{EM}}  & TA & \cite{montminy_improving_2013}\\ 
\cline{3-4}
                                     && \begin{tabular}{@{}c@{}}3-Class \\ DNN\end{tabular}             & \begin{tabular}{@{}c@{}}\cite{carbone_deep_2019}\\ (RSA)\end{tabular} \\
                                     \cline{3-4}
                                     && \begin{tabular}{@{}c@{}}\textbf{256-Class} \\ \textbf{DNN}\end{tabular} & \begin{tabular}{@{}c@{}}\textbf{This Work*} \\ \textbf{(AES-128)}\end{tabular}                               \\

\hline
\end{tabular}
\centering % centering table
\small{\textit{*First EM Cross-Device Deep-Learning Attack on a Symmetric Key Algorithm}}
\end{threeparttable}
\label{tab:litRev}
\end{table}

\section{BACKGROUND \& RELATED WORK}
\subsection{EM Side Channel Attacks}
Since the inception of power SCA ~\cite{kocher_differential_1999}, a wide variety of attacks have been demonstrated, which can be broadly classified into non-profiled attacks like differential/correlational power/EM analysis (DPA, CPA, DEMA, CEMA) ~\cite{brier_correlation_2004}, ~\cite{kocher_differential_1999}, and profiled attacks, such as the statistical template attacks~\cite{chari_template_2003} and ML SCA attacks. While non-profiled attacks perform an attack in a single phase on a target device, profiled attacks consist of two phases, a profiling phase, to learn a leakage pattern and an attack phase, to attack with only a few traces, which practically operate on different devices. During the profiling stage, the attacker will collect traces from a "profiling" device identical to the victim device to build a model. During the attack, this model is then used to recover cryptographic secrets from the victim device. 

\subsection{ML-SCA Attacks}
Template attacks have been shown~\cite{chari_template_2003} to be capable of recovering secret keys with a small number of traces, making them among the most powerful side channel attacks. More recently, supervised ML techniques have been used for profiling SCA~\cite{bartkewitz_efficient_2013}. Among these techniques, DNNs have been one of the most successful, defeating many common countermeasures, such as masking~\cite{gilmore_neural_2015} and clock jitter~\cite{cagli_convolutional_2017}. Table \ref{tab:litRev} provides the summary of related works on profiling attacks. Till date, only one prior work \cite{carbone_deep_2019} has focused on cross-device EM ML SCA attack using only one test device running RSA. Note that this attack required a 3-class DNN~\cite{carbone_deep_2019}, whereas the proposed single-trace (averaged) EM-X-DL attack on AES-128 requires a 256-class DNN, and thus the effects of portability across devices is significantly more prominent. Additionally, AES measurements have significantly lower side-channel SNR compared to a public key algorithm such as RSA. 

\begin{figure}[!t]
  \centering
  \includegraphics[width=0.6\textwidth]{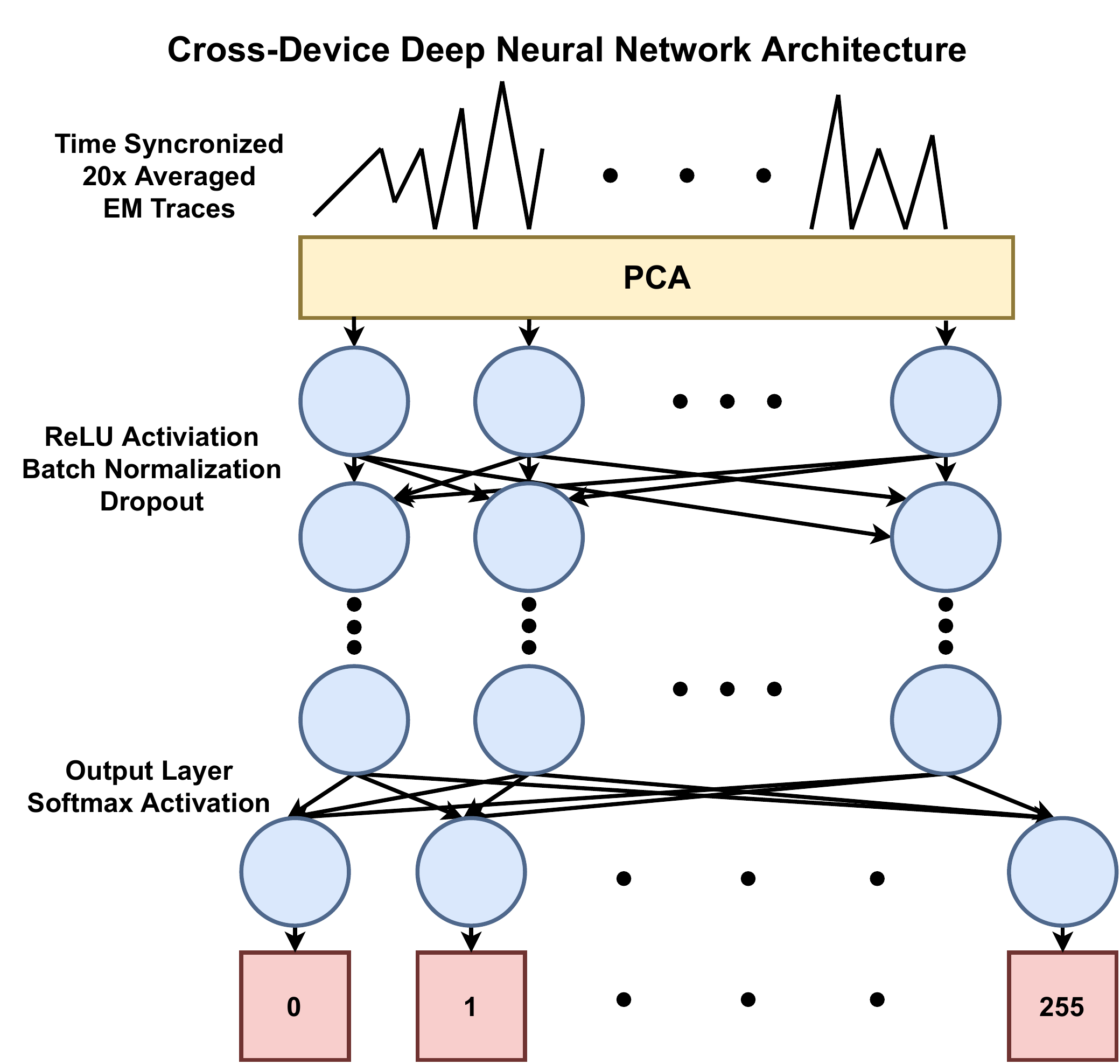}
  \caption{Architecture of the proposed DNN. The network contains 3 dense layers, following each dense layer is a ReLU activation function, batch normalization, and finally a dropout layer. The final output layer provides the output class predictions - the key byte, and thus is size 256, and uses a softmax activation function.}
  \label{nnarch}
\end{figure}

\begin{figure}[!t]
  \centering
  \includegraphics[width=0.7\textwidth]{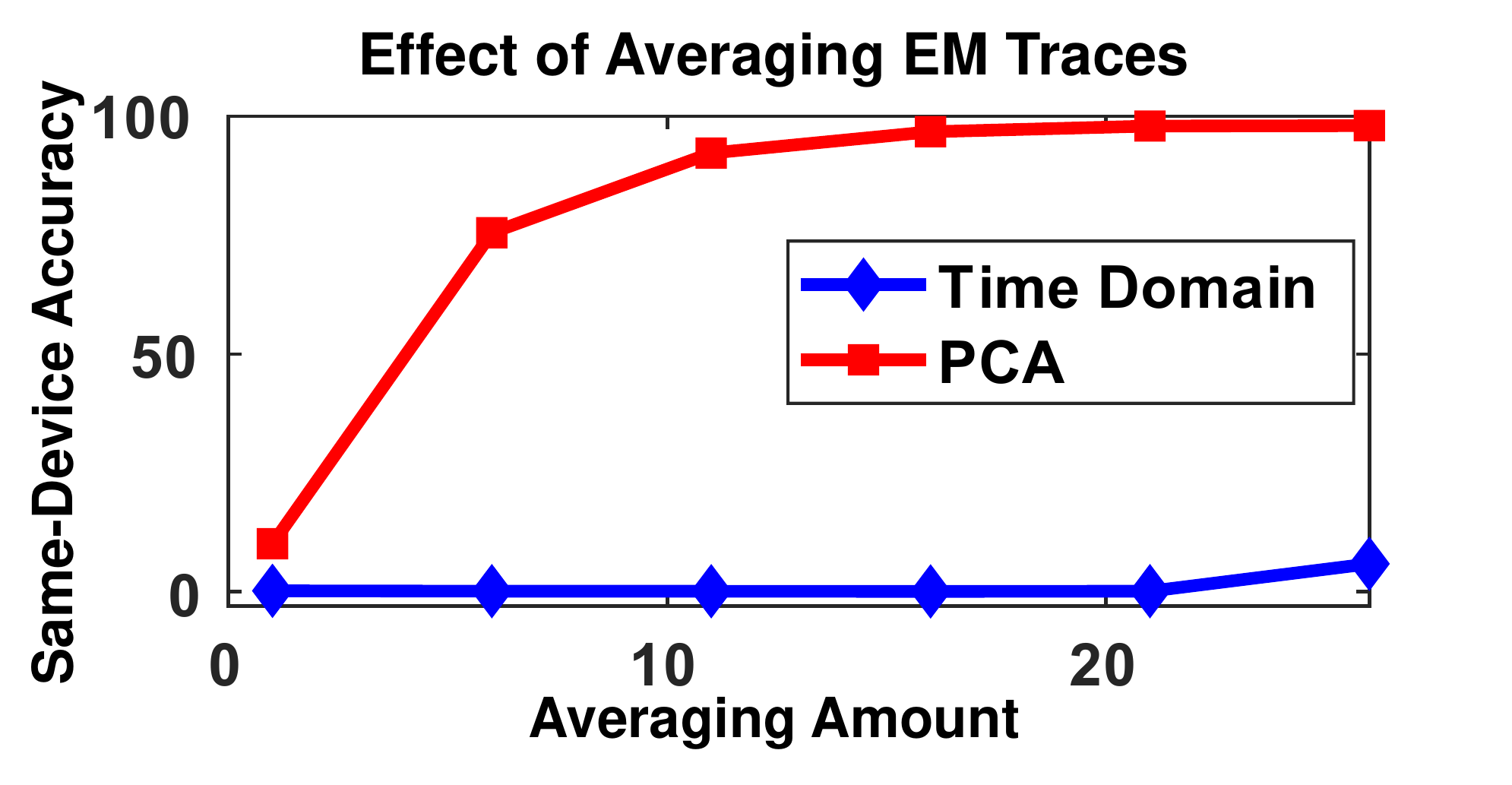}
  \caption{Effect of averaging on the test accuracy of the 256-class DNN when using raw traces and PCA-transformed traces. Increasing averaging hardly allows the DNN to learn from the time-domain EM traces. With PCA used as a pre-processing step, averaging upto $20\times$ smoothly increases the test accuracy to $>99\%$ for the same device.}
  \label{n_avg}
\end{figure}

\section{EM-X-DL SCA ATTACK}
This section evaluates the single-trace (averaged) EM-X-DL attack on AES-128 using a 256-class DNN. For profiling the DNN, EM traces are collected from a set of ten training devices (8-bit Atmega microcontrollers) using the Chipwhisperer~\cite{oflynn_chipwhisperer:_2014} platform, specifically the CW-Lite capture board, along with an off-the-shelf H-field sensor (10mm loop diameter) and a 40dB wideband amplifier. The efficient selection of the training devices is discussed in the subsequent section. For evaluating the attack, ten different devices are reserved separately and the cross-device (EM-X-DL) accuracy is reported as an average of these ten test devices. 

\begin{figure}[!t]
  \centering
  \includegraphics[width=0.75\textwidth]{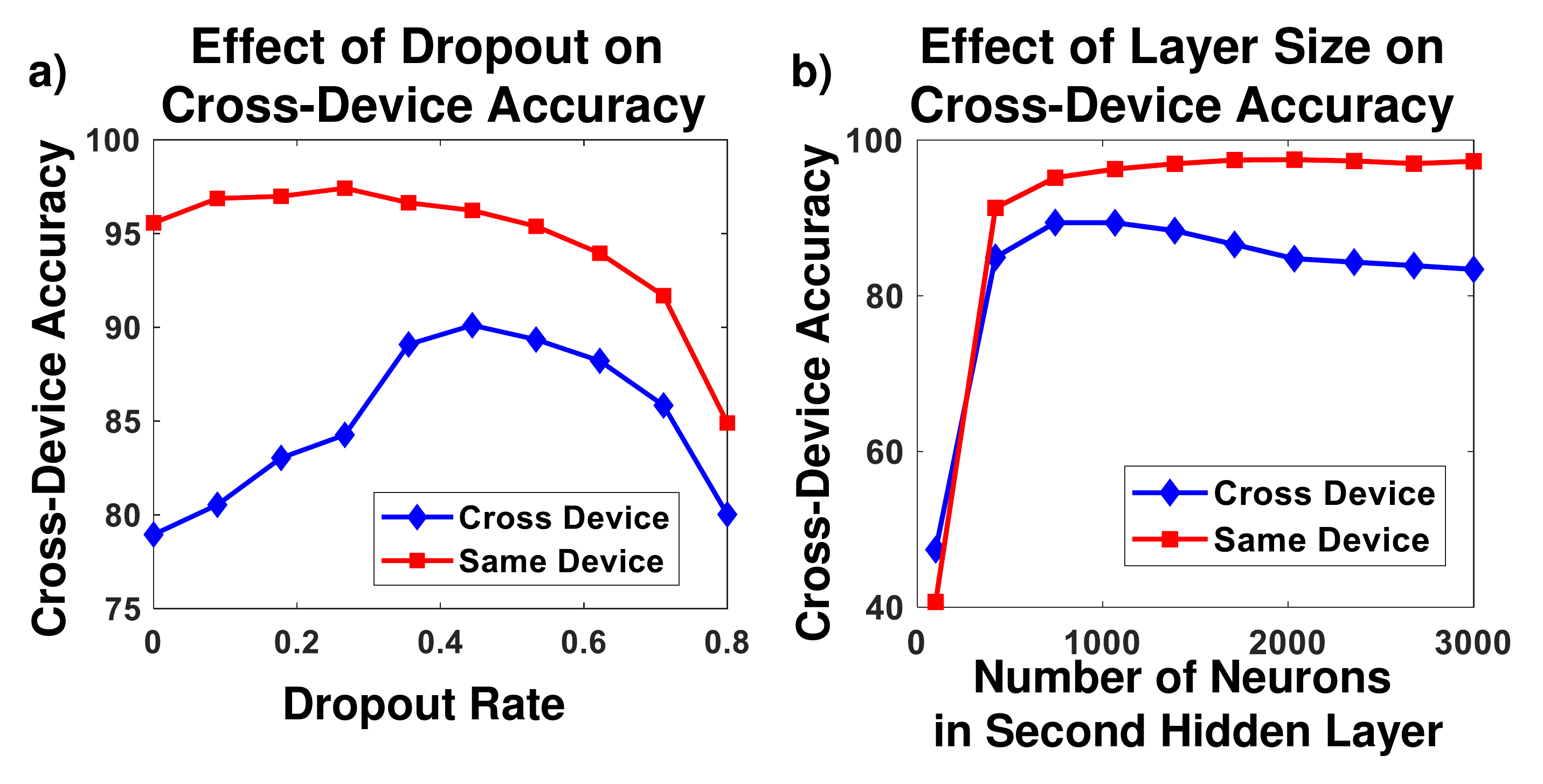}
  \caption{Effect of hyperparameters on both same- and cross-device test accuracy for the PCA-DNN model. (a) Dropout between the first and second hidden layers helps prevent overfitting, maximizing cross-device accuracy at a dropout rate of 0.45. (b) Layer size also demonstrates a similar trend, and reaches maximum cross-device accuracy at $\sim1000$ for the second hidden layer.}
  %Dropout also illustrates the importance of choosing hyperparameters based on cross-device accuracy, rather than the training-device accuracy.
  \label{hyperparam}
\end{figure}

\subsection{Effect of EM Probe Choice}
\begin{figure}[!t]
  \centering
  \includegraphics[width=0.8\textwidth]{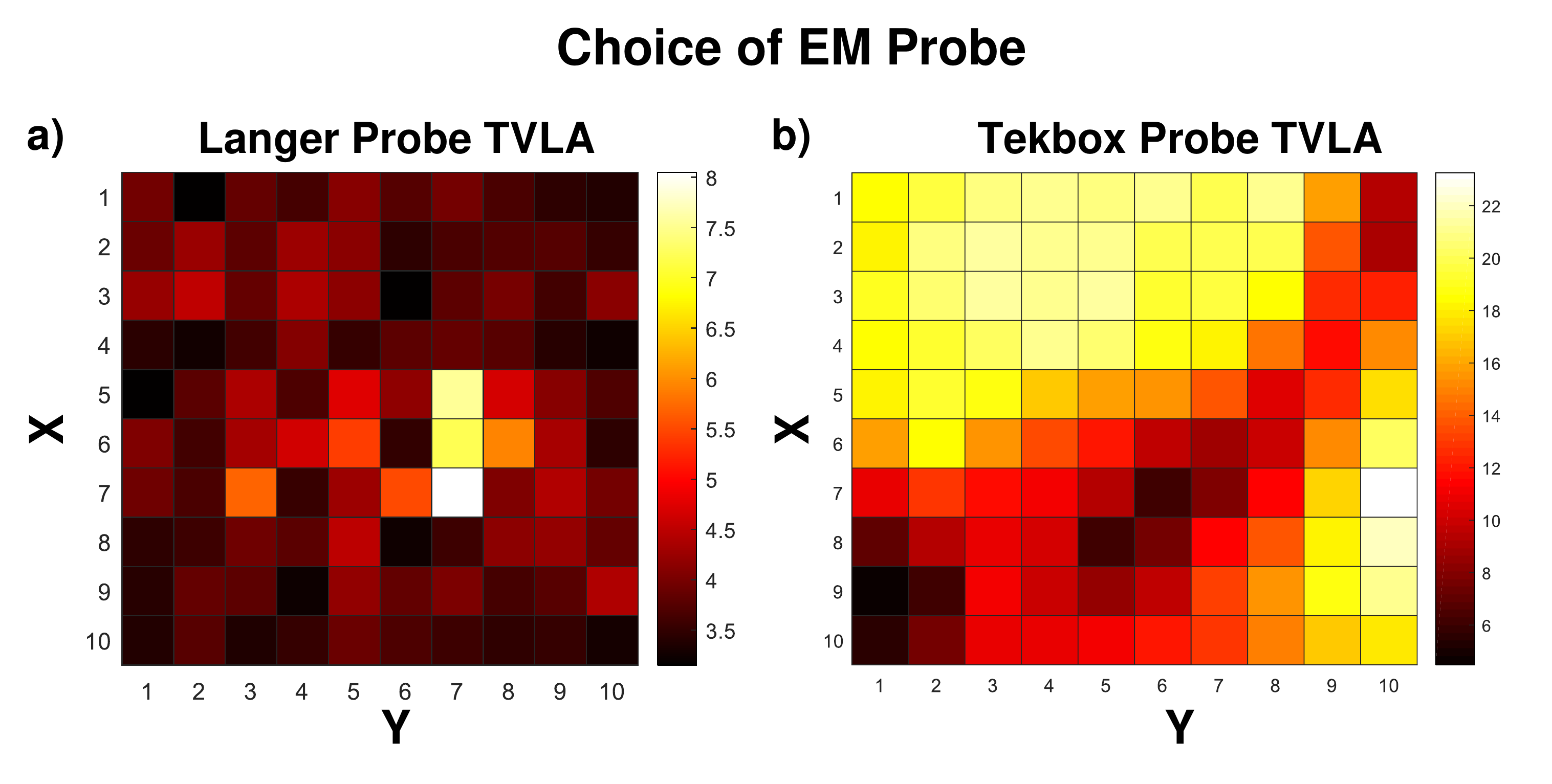}
  \caption{Heatmaps created by performing a fixed vs. random TVLA on a $10 \times 10$ grid spanning the surface of the chip. a) shows results obtained from a Langer ICR HH100-27 probe, while b) shows results from a Tekbox probe. The leakage patterns of the two probes are quite different, which is not unusual as the Langer probe has a much higher spatial resolution. The Langer probe also measured lower side-channel leakage overall, even at the maximum location.}
  \label{probe_compare}
\end{figure}

The EM probe used to collect both training and testing traces has an effect on the side-channel EM signals recorded. Two probes were considered, first a Langer probe with very high spatial sensitivity ($100\mu m$ diameter), and second a texbox probe with low spatial sensitivity ($10mm$ diameter). In order to  estimate the leakage captured by each of the probes, test vector leakage assessment (TVLA)~\cite{becker_test_2013} was used to measure side channel leakage, scanning over the surface of one of the 8-bit X-MEGA devices under attack. The results of these scans can be seen in figure \ref{probe_compare}. As expected, the Langer probe finds high leakage in a very small area, while the larger probe detects leakage over a much larger area of the chip. Additionally, the larger probe detects a much higher level of leakage overall. Since the X-MEGA device is running a software implementation of AES-128, side-channel leakage is not highly localized, as it would be in a hardware implementation, and the high spatial sensitivity of the Langer probe does not provide a large benefit in rejecting algorithmic noise, as there is not a single register at which to target during an attack. For the rest of this work, results will be shown from the larger probe, as the leakage levels are already lower than power, and a variety of effects can be more easily investigated with the relatively higher leakage levels with the larger Tekbox probe - a t-value of 8 on the Langer probe vs. a t-value of 22 with the Tekbox probe, as seen in Figure \ref{probe_compare}.

\subsection{DNN Architecture \& Training}
Figure \ref{nnarch} shows the architecture of the proposed 256-class fully-connected (FC) DNN for the EM-X-DL attack. It should be noted that the EM traces captured using Chipwhisperer are time-synchronized and hence use of a convolutional layer is not necessary~\cite{golder_practical_2019}. 3000 time samples for each trace were collected from the 8-bit microcontrollers running AES-128 clocked at 7.37MHz. 

\begin{figure}[!t]
  \centering
  \includegraphics[width=0.75\textwidth]{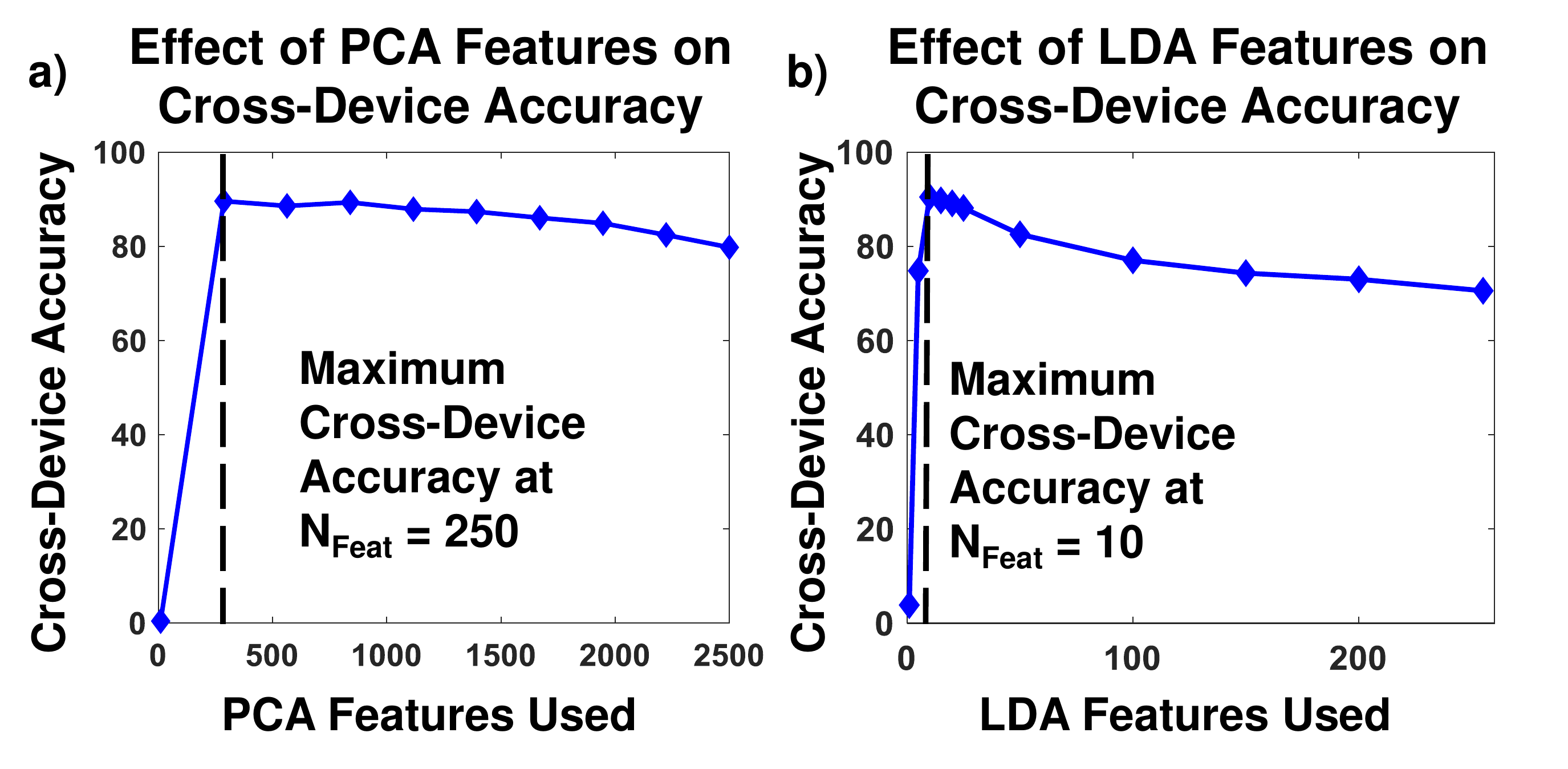}
  \caption{PCA and LDA reach their respective peaks (250 and 10) with relatively few features compared to the size of the original traces (3000). As LDA features are chosen to maximize the class separation, while PCA maximizes variance, LDA is a more efficient technique for this higher dimensional data as it can train the DNN significantly faster. }
  \label{hyperparam_DR}
\end{figure}

\begin{figure}[!t]
  \centering
  \includegraphics[width=0.6\textwidth]{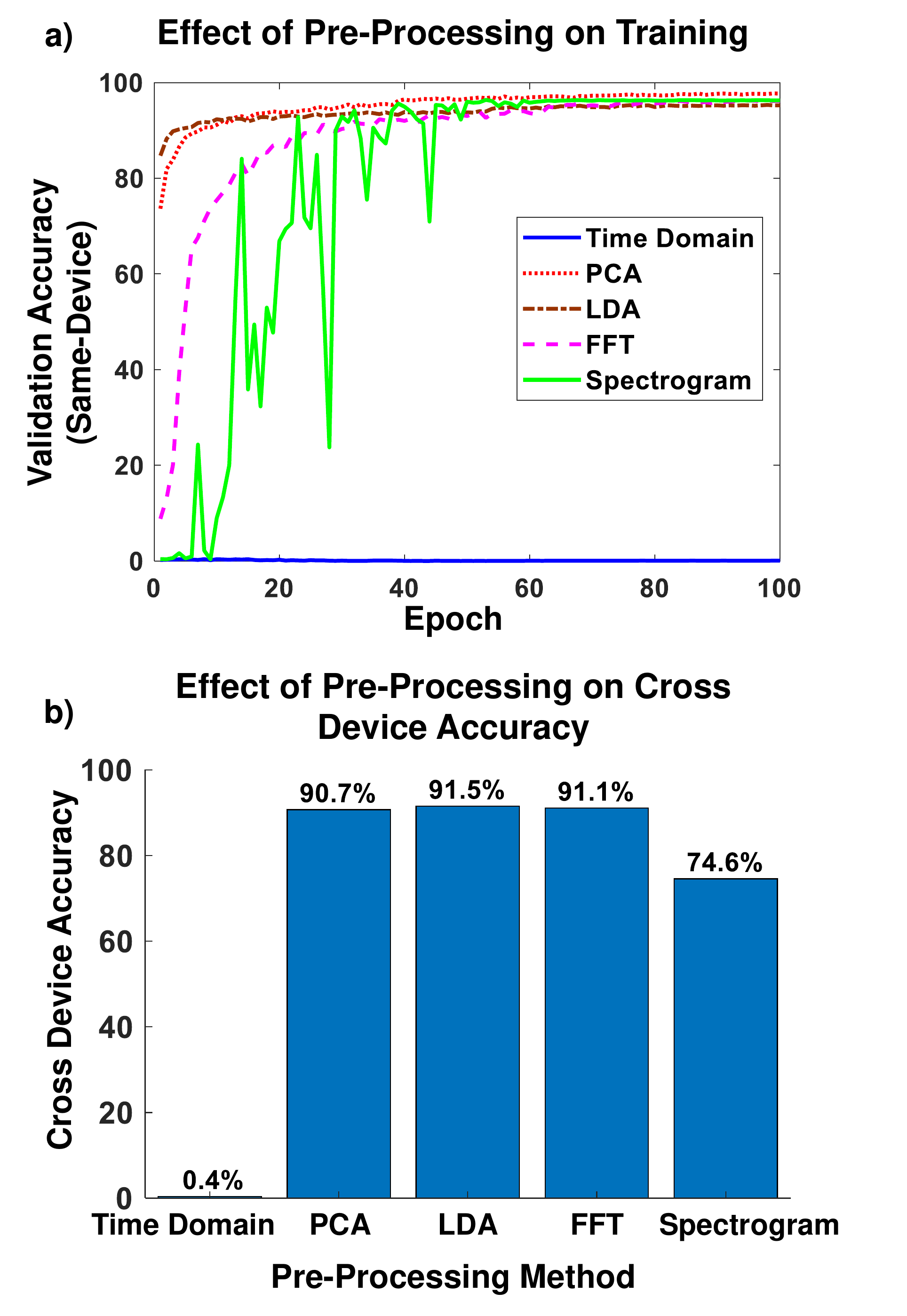}
  \caption{Effect of the different pre-processing techniques on (a) the DNN training accuracy, (b) the cross-device attack (EM-X-DL) accuracy. %All techniques allow the network to train to $>95\%$ accuracy within 100 epochs.
  While all the pre-processing techniques result in high validation (same-device) accuracy, PCA, LDA, FFT result in $>90\%$ cross-device accuracy, while spectrogram yields $74.6\%$ cross-device accuracy.}
  \label{pre_process}
\end{figure}

The DNN, implemented using Tensorflow~\cite{abadi_tensorflow:_nodate}, has a 3000-neuron input layer, followed by three hidden layers with 100, 1024, 512 neurons respectively, and finally the 256-neuron output layer. Rectified Linear Unit (ReLU) activation functions along with batch normalization and dropout used to achieve generalization are utilized for training the DNN. The Adam optimizer, with an initial learning rate of 0.005, which is halved whenever five consecutive training epochs pass without any validation accuracy improvement, is used for training. The effect of different hyperparameters is shown in Figure \ref{hyperparam}. A dropout of 0.45 is the most optimum for the first hidden layer (Figure \ref{hyperparam}(a)), while 1024 hidden neurons for the second hidden layer (Figure \ref{hyperparam}(b)) provides the maximum cross-device accuracy without overfitting to the training devices. For all the results that follow, unless otherwise mentioned, the DNN is trained with ten devices for 100 epochs with a batch size of 64. 

\begin{figure}[!t]
  \centering
  \includegraphics[width=0.65\textwidth]{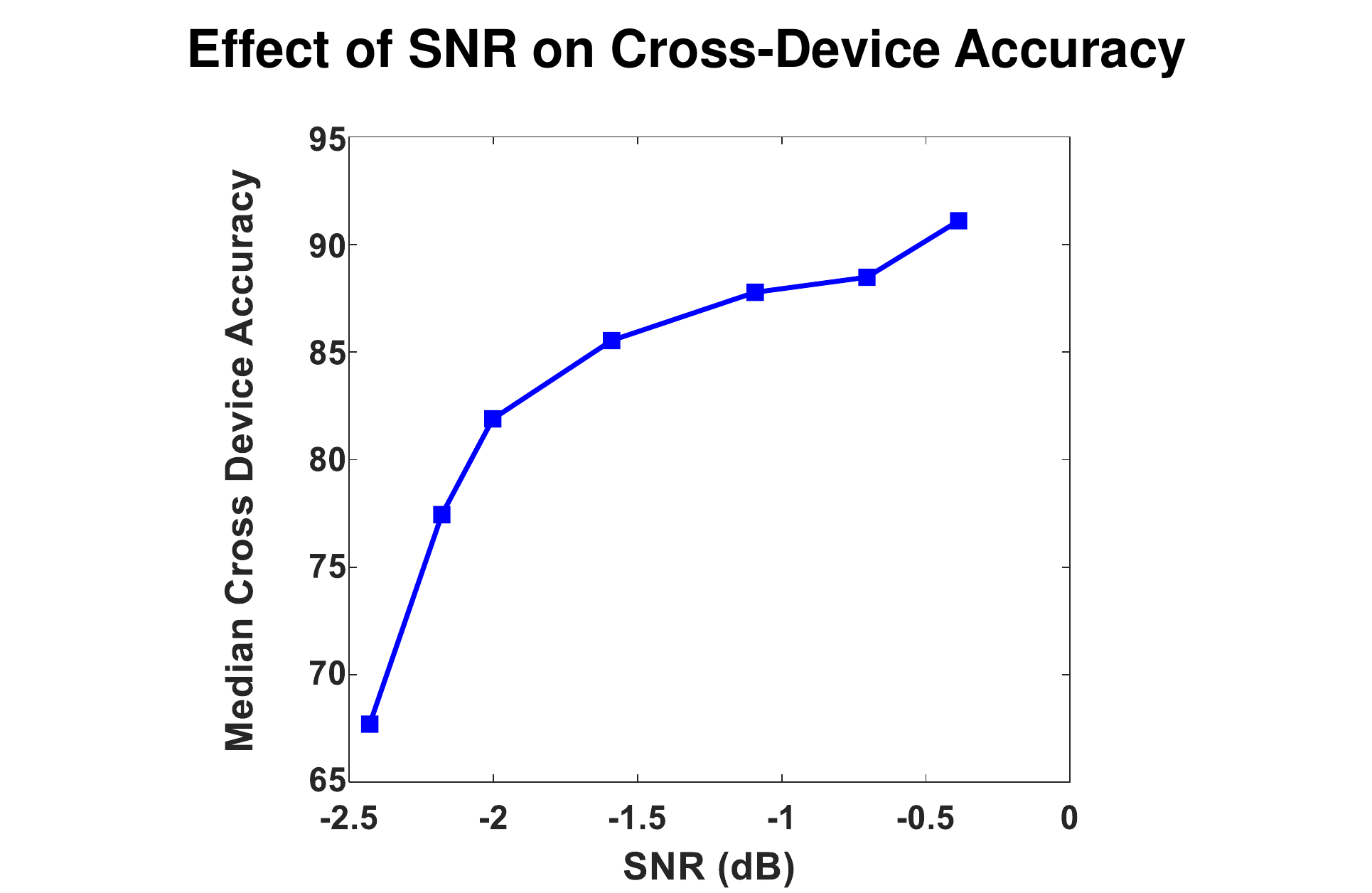}
  \caption{Effect of side-channel SNR on the test accuracy of the 256-class DNN when using PCA-transformed traces. As expected, at higher SNR levels, the DNN achieves higher levels of cross-device accuracy. Note that the limited SNR range is due to the use of avegraging to change SNR, while a majority of SNR reduction is a result of cross-device variations, as seen in Figure \ref{power_vs_em_snr}.}
  \label{acc_snr}
\end{figure}

Now, as the raw EM traces collected from the ten training devices (100K traces each) are fed to the DNN classifier, the validation accuracy remains low ($<1\%$) although training accuracy increases, even after 100 epochs. Figure~\ref{n_avg} (blue curve) shows the effect of averaging on the same-device (test) accuracy. Even with $20\times$ averaging, the time-domain traces shows a test accuracy of $<1\%$, while a dimensionality reduction using PCA achieves $>99\%$ test accuracy for the same device. For cross-device attacks, the accuracy is lower, only 90\% with $20\times$ averaging and PCA. Figure \ref{acc_snr} shows this result in terms of SNR, and shows the DNN's accuracy for lower levels of SNR as well (achieved by lowering the amount of averaging). As expected the accuracy lowers with the SNR, following a similar pattern to Figure \ref{n_avg}. %Hence, averaging combined with pre-processing strategies like PCA achieves a very high test accuracy for the same device. 
Next, we will look into the effect of augmenting traces from ten training devices along with $20\times$ averaging and different pre-processing strategies on the cross-device accuracy. Note that, unless otherwise specified, cross-device accuracy refers to the average key prediction accuracy of the EM-X-DL attack across all the ten test devices.

%The lower SNR of EM traces causes an immediate problem in training ML models. Rather than overfitting to a particular device, ML models trained on raw EM traces both underfit and overfit, as  training accuracy only reaches $50\%$, but validation and test accuracy remain $<1\%$. This suggests the traces do not individually contain enough information to learn from. To address this, we take the common approach of averaging traces with the same key value together before training. The effect of averaging is quite significant, but when using averaged traces directly, accuracy does not improve until $n_{avg} > 25$, and even then is only ~10\%, as seen in Figure~\ref{n_avg}. When principle components analysis (PCA) is used (as discussed in Sec. 3c), the test accuracy smoothly increases from ~10\% to 90\% as averaging increases reaching a maximum when $n_{avg}=20$, again as seen in Figure~\ref{n_avg}. However, eventually, the accuracy decreases again, as the number of averaged traces drops, due to the fact that the number of raw traces collected per device is fixed at 100,000.

\subsection{Single-Trace Attack with Pre-Processing}
In the previous sub-section, it was shown that the averaged time-domain EM traces (100K $\times$ 10 devices) do not train the DNN efficiently, while dimensionality reduction techniques like PCA have a significant impact in training the DNN. Here, we study the effects of PCA~\cite{golder_practical_2019}, LDA ~\cite{renauld_formal_2011} on the time-domain EM traces, as well as the effects of frequency domain based processing (FFT, spectrogram ~\cite{rechberger_practical_2005,yang_convolutional_2019}) on the cross-device accuracy. 

\subsubsection{Dimensionality Reduction using PCA \& LDA}

\begin{figure}[!t]
  \centering
  \includegraphics[width=0.6\textwidth]{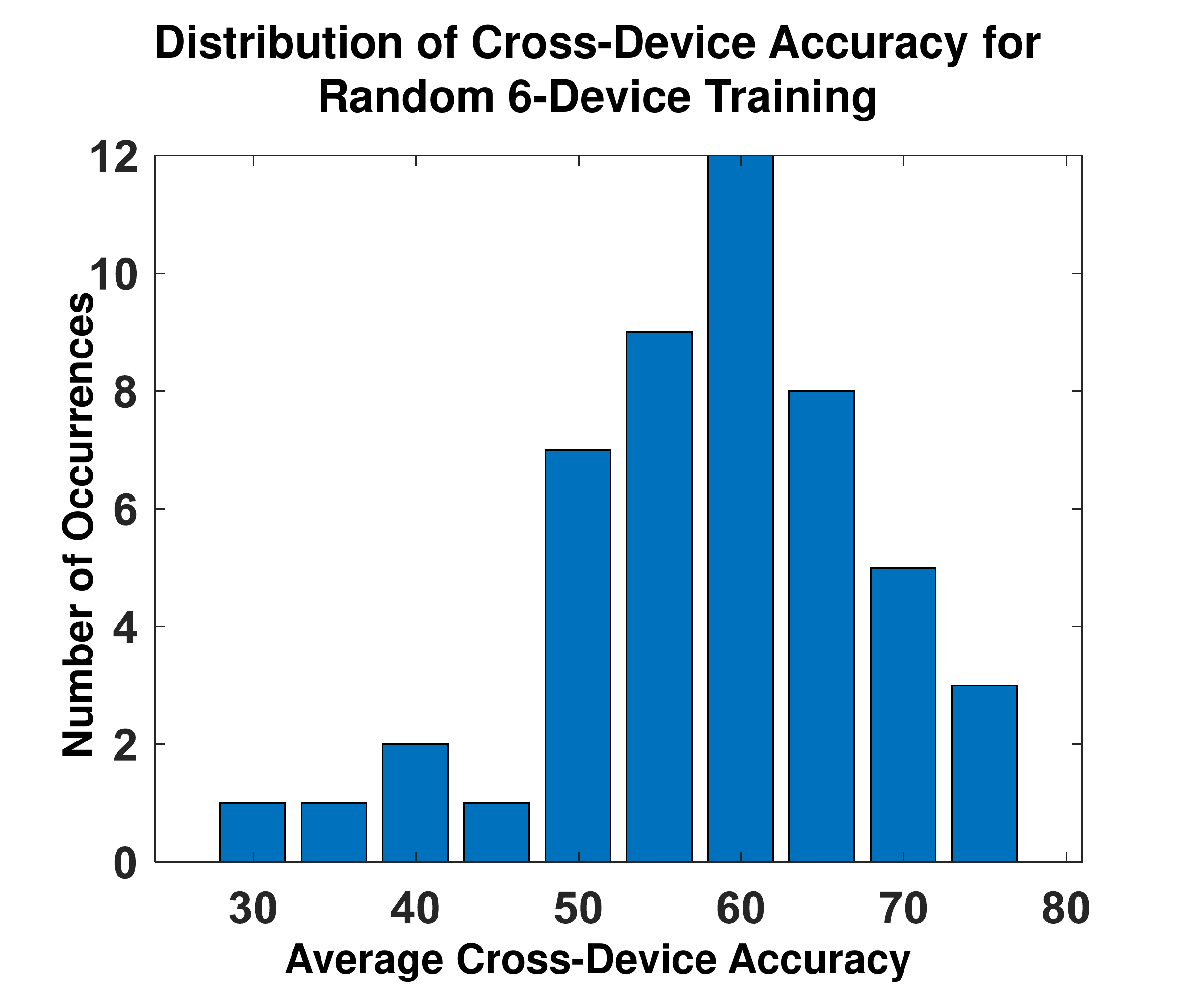}
  \caption{Distribution of cross-device accuracy of the  256-class PCA-DNN trained on random subsets of 6 devices. The mean accuracy is ~60\%, however, the depending on the subset, it can vary significantly between $30 - 75\%$, highlighting the need for an intelligent selection of the training devices.}
  \label{subset_var}
\end{figure}

PCA transforms the input EM trace samples to their principal sub-space where individual features maximize the variance, while LDA achieves the same effect by maximizing the inter-class separation. As seen in Figure \ref{hyperparam_DR}(a, b), the optimal number of features to use in these techniques is much lower than the dimensionality of the raw trace, around 250 in the case of PCA, and a mere 10 in the case of LDA. As shown in Figure \ref{pre_process}(a, b), both of these techniques lead to roughly similar cross-device accuracy, $~91\%$. However, LDA is more efficient as it requires significantly lower training time ($<10\times$) than PCA to achieve the same level of accuracy. 

\subsubsection{Frequency Domain Analysis using FFT \& Spectrogram}
Using FFT on the time-domain averaged ($20\times$) EM traces produces an EM-X-DL attack accuracy of $\sim91\%$ (Figure \ref{pre_process}(b)), which is similar to PCA/LDA. However, it requires higher training time than both PCA and LDA, and hence is not the most efficient approach. Spectrogram combines both time- and frequency-domain information and is naturally two-dimensional. Hence a 2-D CNN ~\cite{simonyan_very_2015} is used for the spectrogram, which achieves a cross-device accuracy of $74.6\%$ (Figure \ref{pre_process}(b)).

\section{EM-X-DL SCA: EFFICIENT SELECTION OF TRAINING DEVICES}

%Algorithm \ref{subset_choice_alg} chooses those particular devices among the entire set (20, in our case - combining all devices) for training, which maximally spans the bivariate distribution and approximates the total PDF of the 20 devices most accurately. 

\begin{algorithm}[!t]
\caption{Algorithm for Device Selection} 
{\textbf{Input:} Trace Samples from all Devices: TraceData, Number of Devices to select: nDev\\
{\textbf{Output:} Subset of size nDev } 
\begin{algorithmic}[!h]
\FOR{$dev = 1 :$ length(TraceData)}
     \STATE $\mu_1 = $ mean(TraceData[dev][:,POI[1])\\
     \STATE $\mu_2 = $ mean(TraceData[dev][:,POI[2])\\
     \STATE meanMap.append({dev, ($\mu_1, \mu_2$)})\\
\ENDFOR
\STATE subset = [1]\\
\FOR{$i=1:$ nDev$-1$}
    \STATE $\mu_{train}$ = mean(meanMap[subset])\\
    \STATE nextDev = $\argmax_j ||\mu_{train} - $meanMap[j][2]$||$
    % \FOR{$j=1:$length(meanMap)}
    %     \STATE dist = $||$curMean - meanMap[j][2]$||$\\
    %     \STATE jmin = 
    %     \IF{dist < minDist}
    %         \STATE minDist = dist\\
    %         \STATE nextDev = meanMap[j][1]\\
    %     \ENDIF
    % \ENDFOR
    \STATE subset.add(meanMap[nextDev][1])\\
    \STATE meanMap.remove(nextDev)\\
\ENDFOR
\STATE return subset
\end{algorithmic}}
\label{subset_choice_alg}
\end{algorithm}

As shown in the previous works \cite{das_x-deepsca:_2019},\cite{bhasin_mind_2019}, the challenge of a ML SCA model being able to accurately classify traces collected from devices it has not been trained with, can be addressed by training with a variety of devices, so that the model does not overfit to the particular leakage pattern of one device. This remains true when using EM traces, however, many more devices are required to gain a high level of cross-device accuracy, because, as seen in Figure \ref{power_vs_em_snr}, EM measurements are more sensitive to cross-device variations. Moreover, averaging clearly plays a key role, further increasing the number of traces required. Thus, it is of interest to be able to train using the smallest possible set of devices, reducing both the number of traces needed as well as the training time for the DNN. For this, two things must hold true: \textit{First}, the choice of devices must affect the cross-device accuracy for a given number of devices, and \textit{second}, there must be a way of determining whether or not to include a device for training from a small sample of traces.

\subsection{Cross-Device Accuracy Variance}
To address the first point, the effect of the subset, the EM-X-DL model is trained with a random subset of six devices, then tested against all the remaining fourteen devices. As shown in Figure~\ref{subset_var}, the average cross-device accuracy can vary greatly even for a set of only six devices, with accuracy ranging from 10\% to 75\% for different six-device combinations. This shows that there are \textbf{subsets of training devices that can improve accuracy rather than simply adding more devices.} However, as there are a large number of possible subsets for a given size, an algorithm is necessary to choose one such subset which results in high cross-device accuracy. Such an algorithm would then enable an attacker to gather quick measurements from a large set of devices, and determine a small subset of devices to collect a large number of traces from, for training the DNN model.

\begin{figure}[!t]
  \centering
  \includegraphics[width=0.6\textwidth]{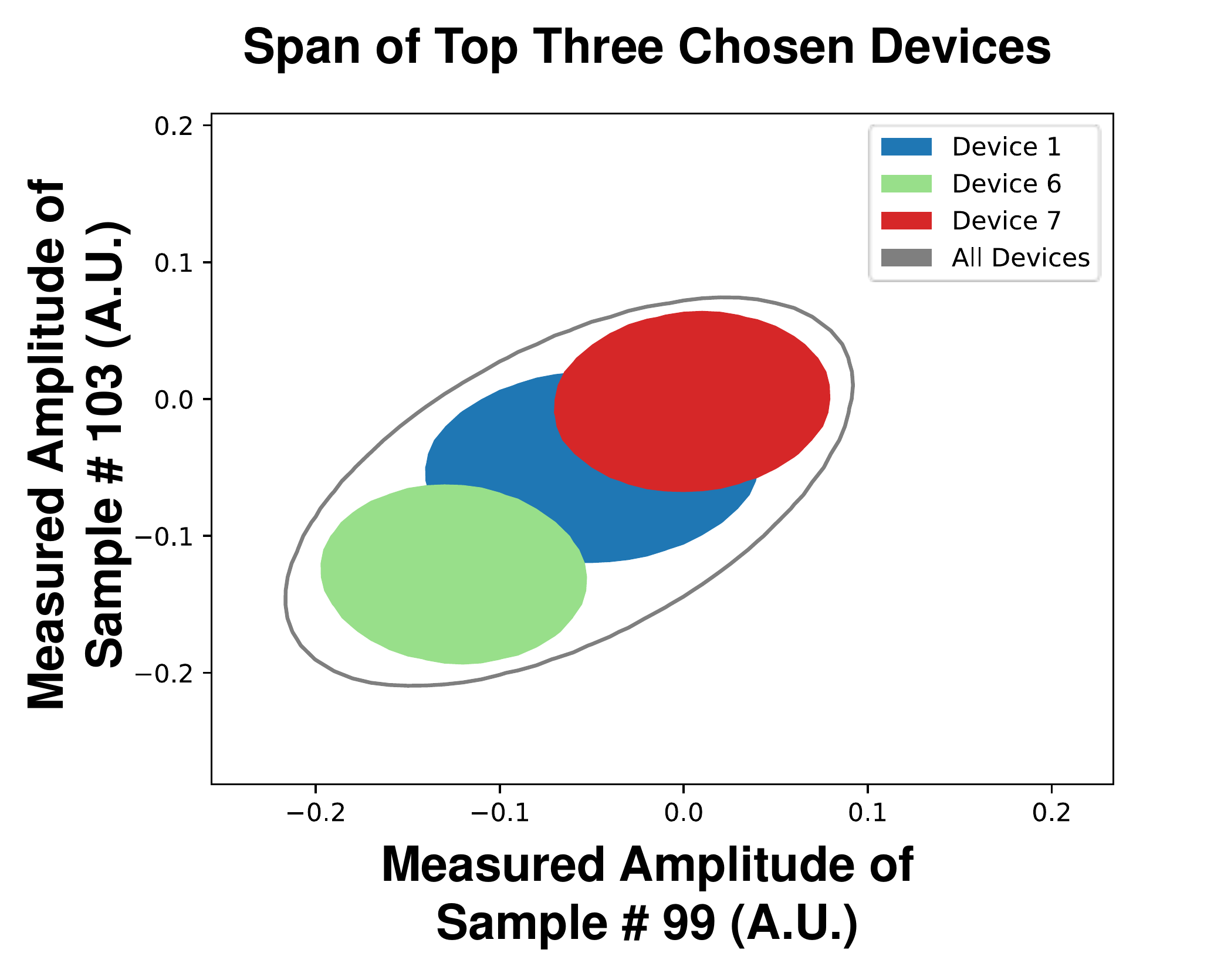}  \caption{Bivariate analysis of the first 3 devices chosen by Algorithm \ref{subset_choice_alg}. The top three chosen devices already span a large portion of the distribution containing all devices.}
  \label{3_dev_dist}
\end{figure}

\subsection{Bivariate POI Based Device Selection}
The proposed algorithm begins by identifying two points of interest (POIs) in the traces. This can be done through any POI identification technique, here POIs are chosen as time samples which have the highest difference of means (DOM). Once the top two POIs are found, the mean $\boldsymbol{\mu_i} = (\mu_{POI1}, \mu_{POI2})$ of this POI pair is calculated across all traces for each device. Then, to construct the subset of devices for training, one device is initially chosen arbitrarily, and additional devices are added as follows: The mean POI pair of all devices currently included in the training subset, $\boldsymbol{\mu_{train}}$ is calculated. Then, the next device is chosen such that $||\boldsymbol{\mu_i} - \boldsymbol{\mu_{train}}||_2$ is maximized, where $i$ varies over all devices not already included in the training subset. In this way, at each step, the device whose top two average POIs are furthest from the average POIs of the currently selected devices is added to the training set. This method is detailed in Algorithm \ref{subset_choice_alg}.

\begin{figure}[!t]
  \centering
  \includegraphics[width=0.6\textwidth]{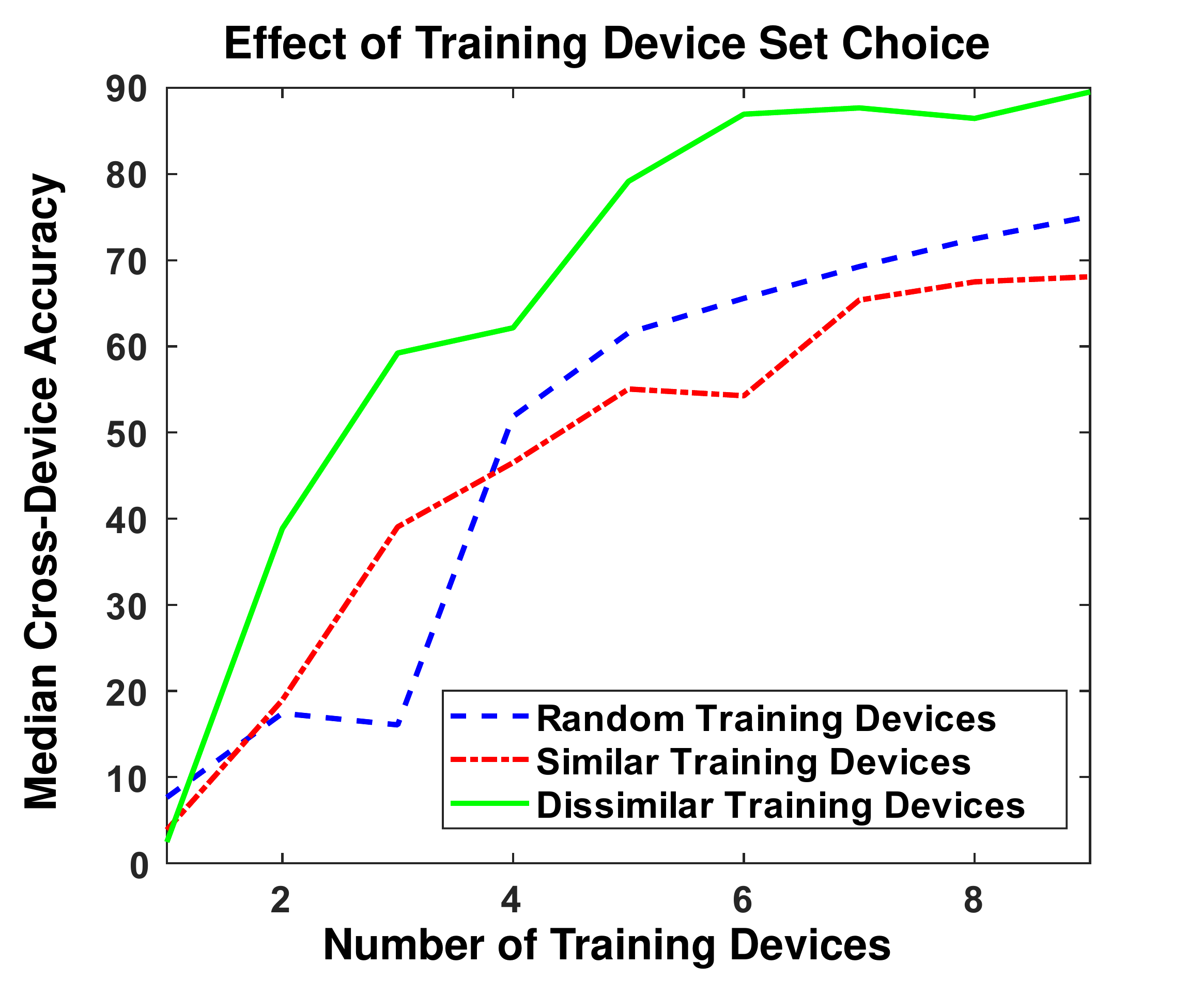}
  \caption{Depending on the choice of devices used for training, cross-device accuracy varies significantly. Choosing ``dissimilar" devices by algorithm \ref{subset_choice_alg} gives high accuracy, while choosing ``similar" training devices yields a low cross-device accuracy. Randomly selecting devices shows slightly higher test accuracies than choosing ``similar" devices. }
  \label{subset_choice} 
\end{figure}

\begin{figure*}[!t]
  \centering
  \includegraphics[width=1\textwidth]{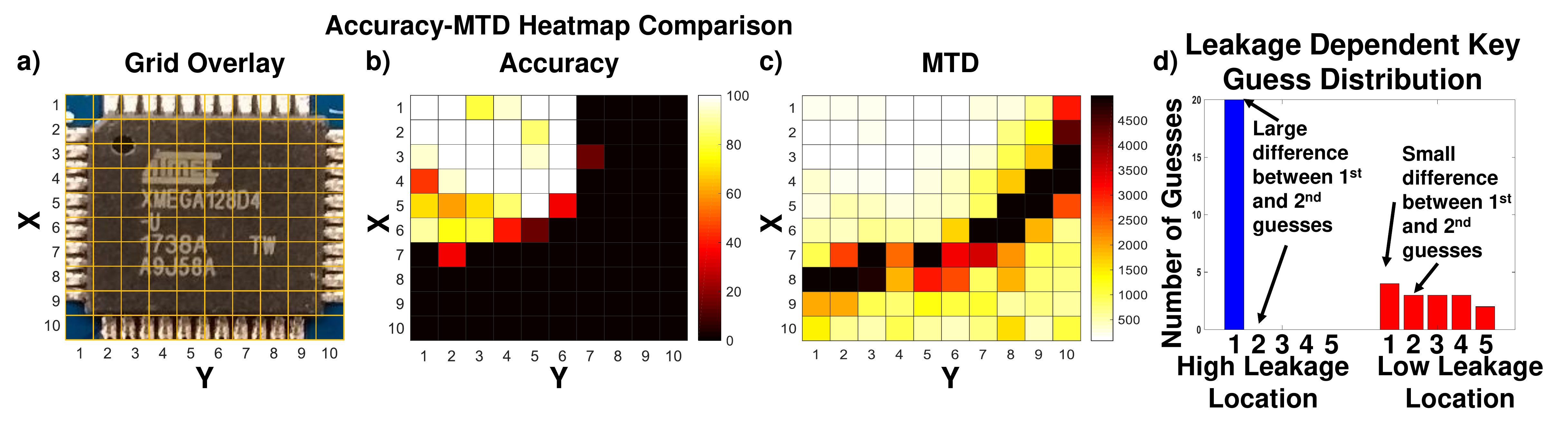}
  \caption{(a) $10\times10$ virtual grid overlay of the chip. (b, c) Comparison of EM-X-DL model accuracy to CEMA-MTD. The ML model is able to predict with high accuracy in the region of the chip with low MTD values, however, when the MTD rises above 250, the model is unable to correctly predict the key values. (d) EM-X-DL model predictions on 20 samples from a high leakage location (1,1), and a low leakage location (9, 4) on a test device. At a location with high leakage, the frequency of the highest predicted key byte value is distinguishable from the next, demonstrating the high confidence of the attacker.}
  \label{heatmap_compare}
\end{figure*}
Figure \ref{3_dev_dist} shows the 2-D bivariate normal distribution of the first three devices chosen using this algorithm, along with the total distribution of all devices. With these three devices, a large portion of the distribution spanned by all the devices is covered, revealing the successful operation of the algorithm. Importantly, this algorithm also provides the desired results during training, shown in Figure \ref{subset_choice}, as using this algorithm to choose the training devices gives higher cross-device (EM-X-DL) accuracy for any number of devices. Additionally, training with the devices closest to the current training set, as opposed to the furthest away, results in cross-device accuracy significantly lower than the maximally different devices, and generally lower accuracy than randomly selected devices as well. These results were obtained with the proposed 256-class DNN, using $20\times$ averaging and PCA-based pre-processing. From Figure \ref{subset_choice}, we also see that to attain a certain cross-device accuracy, this algorithm requires between $20\% - 40\%$ fewer training devices compared to random device selection.

\section{EM LEAKAGE ASSESSMENT \& ATTACK}
Once the DNN model for the EM-X-DL SCA is trained, the main goal of an attacker is to break the secret key with minimum number of traces from an identical but unseen target device. This section demonstrates an end-to-end attack strategy using the EM-X-DL model on a new device. By scanning the surface of the victim microcontroller and collecting traces at each point (seen in Figure \ref{heatmap_compare}(a)), the heatmap in Figure \ref{heatmap_compare}(b) was created by classifying the traces and determining the test accuracy for each point. As all training traces were collected from the same location (with maximum leakage on the chip evaluated using test vector leakage assessment (TVLA)), as expected the accuracy is highest in this region, then drops off sharply further from the measurement point. Figure \ref{heatmap_compare}(c) shows the minimum traces to disclosure (MTD) from a CEMA attack over the same chip. Comparing this to the accuracy heatmap shows that the ML model can correctly classify traces that are collected from a location which has an MTD less than $250$.

Now, in this virtual grid, to converge to the best location for the EM-X-DL attack on the new device, the attacker can query the EM-X-DL model with multiple averaged traces collected from the test device and \textbf{observe if the frequency of the highest predicted key byte is distinguishable from the next.} Should leakage be present, the correct key byte would be predicted more often than others. If leakage is not present, predictions would be split between several key values. Thus, the ratio between the first and second most commonly predicted value provides a measure of the attacker's confidence in the prediction. This effect is shown in Figure \ref{heatmap_compare}(d), which shows the five most common predictions for both a location of high leakage,(1,2) (left) and low leakage, (2,9) (right). Note that, with this prior knowledge of the heatmap, the attacker can also divide the chip into 4 quadrants (for this particular chip) and get the correct key from the left most quadrant with a very high confidence. 

\begin{table}[!t]
% centering table
\centering
\begin{tabular}{@{}cccc@{}}
\hline % inserting double-line
\\ [-1ex]
% $\boldsymbol{Preprocessing}$ & \multicolumn{3}{c}{$\boldsymbol{Cross-Device \ Accuracy \  (\%)}$}\\
${Preprocessing}$ & \multicolumn{3}{c}{${Cross-Device \ Accuracy \  (\%)}$}\\
${Technique}$  & Minimum & Average & Maximum
\\ [1ex]
\hline % inserts single horizontal line
\\ [-1ex]
Time Domain &0.28  &0.37  &0.45 \\ % inserting body of the table
PCA &81.27  &90.72  &96.77\\
LDA &81.21  &91.52  &96.42 \\
FFT &82.40  &91.07  &95.50 \\
Spectrogram &30.53  &74.58  &94.02 \\  [1ex]
\hline 
\end{tabular}
\caption{Cross-Device attack Performance of Deep Learning-based Methods for different Pre-Processing Techniques} % title name of the table
\label{tab:PPer}
\end{table}

\section{REMARKS \& CONCLUSION}
This work showed a Cross-device Deep Learning based EM (EM-X-DL) SCA attack on a symmetric key encryption engine (AES-128) in a low SNR setting. Utilizing a 256-class DNN, averaged EM traces from 10 training devices along with dimensionality-reduction based pre-processing (like LDA) the model achieves $\sim91.5\%$ EM-X-DL single-trace (averaged) attack accuracy against another set of ten test devices. Table~\ref{tab:PPer} summarizes the EM-X-DL attack accuracy for each of the different techniques studied in this article. An algorithm for efficient selection of training devices is proposed to speed up the profiling phase. Finally, an end-to-end attack using EM scanning is demonstrated showing that the attacker can detect the position of highest leakage on the chip using the proposed EM-X-DL model along with the secret key with high confidence.  

For the future scope of this work, the end-to-end EM-X-DL attack can be more generalized by capturing traces from multiple locations across the chip, rather than a single location, for training the DNN. This would make the EM-X-DL attack much more efficient and faster as the attacker would be able to extract the key without having to detect one of the highest leakage locations on the chip.

Additionally, it was shown that the SNR of traces used for training and testing have a strong impact on the accuracy of the produced DNN as expected. This encourages development of countermeasures focused on reducing the SNR of side-channel signals, such as \cite{das_stellar:_2018}. While such countermeasures can always fundamentally be defeated by collecting additional traces, by reducing the SNR significantly, collecting a sufficient number of traces from a large enough variety of devices becomes infeasible.

%Continuation of this work would involve investigating cross-device leakage assessment, and finding a relationship between MTD via CEMA and key byte classification accuracy. Determining the effect of trace collection position on the efficacy of EM-X-DL would be an additional direction for future work. For instance, by collecting traces from locations distributed across the surface of the training devices, rather than only a single location, ML models may be able to generalize more effectively.

%\section{Conclusion}
%This paper has demonstrated the first cross-device EM-ML-SCA attack. By using the concept of multi-device training used in cross-device Power-ML-SCA attacks, along with additional pre-processing, a model with $>90\%$ single-trace accuracy was developed, summarized in Table~\ref{tab:PPer}. In order to offset the increased requirement on the number of devices needed for training, an algorithm for choosing devices based on a small number of measurements was created, and demonstrated to provide maximum cross-device accuracy for any number of training devices used. Finally, the concept of using the proposed model to measure EM leakage was proposed, and initial results were presented.
\bibliographystyle{ACM-Reference-Format}
\bibliography{DAC_2020}
\end{document}